\documentclass[aps,prb,twocolumn,showpacs]{revtex4}
\usepackage{bm}
\usepackage{graphicx}
\usepackage{amsmath}
\usepackage{eufrak}
\usepackage{color}
\usepackage{subfigure} 
\usepackage[colorlinks=true]{hyperref} 
\newcommand{\nix}[1]{}

\begin{document}

\title{Quantum Oscillations of Photocurrents in HgTe Quantum Wells with Dirac and Parabolic Dispersions}
 
\author{C.\,Zoth,$^1$ P.\,Olbrich,$^1$  P.\,Vierling,$^1$ K.-M.\,Dantscher,$^1$ 
V.V.\,Bel'kov,$^2$ M.A.\,Semina,$^2$ M.M.\,Glazov,$^2$ 
L.E.\,Golub,$^2$ D.A.\,Kozlov,$^{3,4}$  Z.D.\,Kvon,$^{3,4}$  
N.N.\,Mikhailov,$^3$ S.A.\,Dvoretsky,$^3$ 
and S.D.\,Ganichev$^1$}
\affiliation{$^1$Terahertz Center, University of 
Regensburg, 93040 Regensburg, Germany}
\affiliation{$^2$Ioffe Physical-Technical Institute, Russian
Academy of Sciences, 194021 St.\,Petersburg, Russia}
\affiliation{$^3$A.V. Rzhanov Institute of
Semiconductor Physics, Novosibirsk 630090, Russia}
\affiliation{$^4$Novosibirsk State University,
Novosibirsk, 630090, Russia}

\begin{abstract}
We report on the observation of magneto-oscillations 
of terahertz radiation induced photocurrent  in HgTe/HgCdTe quantum wells 
(QWs)  of different widths, 
which are characterized by a Dirac-like, inverted and normal parabolic 
band structure. The photocurrent data are accompanied by measurements 
of photoresistance (photoconductivity), radiation transmission, as well as 
magneto-transport. We develop a microscopic model of a cyclotron-resonance 
assisted photogalvanic effect, 
which describes main experimental findings. We demonstrate that the quantum oscillations 
of the photocurrent 
are caused by the crossing of Fermi level by Landau levels resulting in the 
oscillations of spin  polarization and electron mobilities in spin subbands. 
Theory explains a photocurrent direction reversal with the variation of magnetic field observed in experiment. We describe the photoconductivity oscillations related with the thermal suppression of the Shubnikov-de Haas effect.
\end{abstract}

\pacs{}
\maketitle

\section{Introduction}

The physics of relativistic Dirac fermions in semiconductors has became a topical 
field of condensed matter due to their unique electronic, optic and optoelectronic properties. 
One can distinguish two groups of such materials, characterized by either weak spin-orbit coupling 
like graphene, for recent reviews see~Refs.~\onlinecite{CastroNeto2009,Novoselov2012,Glazov2014}, or by rather strong spin-orbit interaction, typical for the most of topological insulators, 
for reviews see~Refs.~\onlinecite{Hasan2010,Moore2010,Zhang2011}.
Among the representatives of the latter group, the HgTe-based crystalline structures have attracted 
particular attention, because they allow one to fabricate two- and three-dimensional topological insulators.\cite{Bernevig2006,Koenig2007,Roth2009,Moler2013,Kvon2014,Fu2007,Dai2008,Bruene2011,Oostinga2013,Kozlov2014}
In this very system, one can obtain  a Dirac-like, inverted and normal energy dispersions
without changing the material.\cite{Bernevig2006,Koenig2007,Dyakonov1981,Kvon2008,Diehl2009,Wittman2010,Ikonnikov2010,Buettner2011,Kvon2011,Hancock2011,Ikonnikov2011,Zholudev2012,Ikonnikov2012,Kozlov2012,Olbrich2013,Shuvaev2013,Shuvaev2013a,Orlita2014} Thus, investigating various electronic properties in 
HgTe-based quantum wells with different thicknesses one can address similarities and differences 
in phenomena excited for different types of electron energy spectra.

Here we report on the complex study of photocurrent, photoresistance, optical transmission, and 
electron transport  in HgTe quantum wells with the thicknesses ranging from 5 to 21~nm where 
possible variants of energy spectrum are realized.
While the terahertz (THz) radiation  induced  photogalvanic 
currents~\cite{Ganichev02,Ivchenkobook,Ganichevbook,Dyakonov2008} in HgTe QWs subjected to 
a classical
magnetic field $\bm B$ are studied in details in Refs.~\onlinecite{Diehl2009,Olbrich2013}, 
our paper focuses on the observation and analysis of quantum oscillations in the 
cyclotron-resonance-assisted photocurrent excited 
by THz laser radiation. 
We show that the photocurrent quantum oscillations, similar to the  de Haas-van Alphen 
and Shubnikov-de Haas effects, stem from the consecutive crossings of Fermi level by Landau levels, 
but are drastically enhanced due to the cyclotron resonance (CR).  We discuss the microscopic origin of the photocurrent 
in all three cases of electron energy dispersion and demonstrate that it is caused by the magnetogyrotropic 
photogalvanic effect~\cite{Belkov2008}. 
While the main features of the phenomena, such as $1/B$-periodic oscillations 
superimposed by the CR resonance, are very general and the effect is of the same order of magnitude for all studied 
samples, strong peculiarities for Dirac fermions have been observed. Particularly, as a distinguishing 
feature of the linear spectrum, cyclotron resonance and quantum oscillations in the photocurrent 
are obtained simply by the variation of the carrier density in a QW.

\section{Samples, magneto-transport data and methods}

\subsection{Samples}

The experiments are carried out on doped (013)-oriented MBE grown 
Hg$_{0.3}$Cd$_{0.7}$Te/HgTe/Hg$_{0.3}$Cd$_{0.7}$Te single QW structures~\cite{Kvon2009} with 
different widths, $L_{w}$, of 5, 6.6, 8, and 21\,nm, mobilities of about 
$10^5$\,cm$^2$/{(V{$\cdot$s)} at $T = 4.2$~K and carrier densities $n$ in the range of 
$5 \times 10^{10} \div 7.5 \times 10^{11}$~cm$^{-2}$.
In HgTe, an increase of the QW thickness results in the qualitative change 
of the band
structure,\cite{Bernevig2006,Koenig2007} starting with a normal parabolic dispersion 
(5~nm QW), switching to Dirac cones (6.6~nm) and then to an inverted, close
to parabolic, band structure (8 and 21\,nm).
Besides the structures with pure HgTe QWs, we also studied 
Hg$_{0.86}$Cd$_{0.14}$Te QWs with the same barriers but containing 14\% Cd in the QW layer. 
The most important difference of such QWs, compared to that made of pure 
HgTe, is that the transition from a normal to an inverted
energy spectrum is observed for wider QWs.\cite{Ikonnikov2010,Ikonnikov2012}
This fact allows us to study the same phenomena in material with inverted or 
non-inverted band structures in QWs of the same thickness, here $L_w$ = 8~nm.

The samples are prepared in different geometries including Hall bar
structures, without gate and with a semitransparent gate, as well
as square-shaped samples of 5$\times$5 mm$^2$ size. While the square-shaped 
large-size structures require van der Pauw geometry for transport measurements, 
they are prepared in order to enable simultaneous
measurements of the photoresponse and radiation transmission. The
typical structure designs and the ohmic contacts positions for
Hall-bar and square-shaped samples are shown in
Fig.~\ref{fig_0}. Note, that for the
square-shaped samples eight ohmic contacts have been prepared in
the middle of the edges and on the corners of the structure. For
magneto-optic and magneto-transport experiments a magnetic field
$\bm B$ up to 7\,T is applied normal to the QW plane.

\begin{figure}[t]
\includegraphics[width=0.7\linewidth]{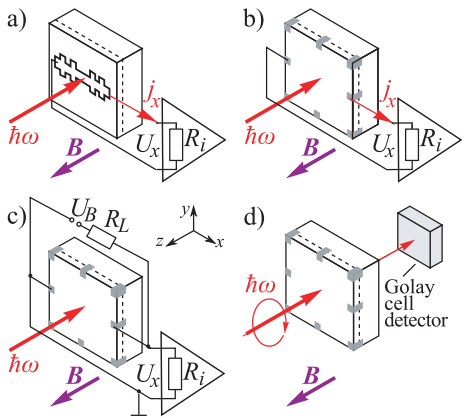}
\caption{Experimental geometries of
photocurrent measurements  for Hall bar- and square-shaped 
samples in a) and  b), respectively, for c) photoresistance, 
as well as d) transmission.}
\label{fig_0}
\end{figure}

\subsection{Magneto-transport data}

In all samples well-pronounced Shubnikov-de Haas (SdH)
oscillations of static conductivity, see dashed lines in Figs.~\ref{fig_1} to
\ref{fig_6}, and quantum Hall plateaus (not shown) have been
detected. Magneto-transport has been measured applying slowly modulated bias 
($f$ = 12 Hz, 1 V) to the sample. The carrier densities $n$ at 4.2~K 

are given in Table~\ref{tab_1}.

At \textit{high} magnetic field, SdH oscillations
corresponding to both, $even$ and $odd$ filling factors, and having similar amplitude are observed
in all samples characterized by almost parabolic dispersion, see Figs.~\ref{fig_1} to
\ref{fig_4}. This indicates that the absolute value of the Zeeman splitting
$|\Delta_Z|$ is comparable to the energy difference between neighbouring levels $\Delta E = \hbar\omega_c - |\Delta_Z|$. 
Here $\Delta_Z = g\mu_B B$ with $g$ being the effective
electron $g$-factor, $\mu_B$ being the Bohr magnetron, $B = |\bm B|$ being 
the magnetic field directed along the sample normal, the cyclotron frequency $\omega_c = |e B|/m_c c$, 
with carrier  charge $e$, speed of light $c$, and the cyclotron mass $m_c$. 
The latter is given by $m_c = \hbar^2kdk/dE,$ where $k$ and $E=E(k)$ are wavevector 
and energy, respectively. In the particular case of linear energy dispersion 
characterized by a constant velocity $v_{\rm{DF}}$ the cyclotron mass $m_c$ depends 
 on the Fermi energy $E_{\rm F}\,$=$\,\sqrt{2\pi n}(\hbar v_{\rm{DF}}$) as $m_c = E_{\rm F} / v^{2}_{\rm DF}$.\cite{briskot}

Measuring the SdH oscillations in samples with parabolic
dispersion subjected to \textit{low} magnetic fields we detect either 
$even$ or $odd$ numbers of minima, depending on the electron 
density of the sample. 
This is caused by the fact that at low magnetic fields the 
distance between neighboring levels $\Delta E$ 
is smaller than the level broadening caused by the electron scattering processes. 
Hence, $even$ minima are observed if the absolute value of the Zeeman 
splitting $|\Delta_Z|$ is smaller than $\Delta E$.
Such a behavior is detected, e.g., for 
magnetic fields $B \lesssim 2.5$~T in the 
Hall bar 8~nm Hg$_{0.86}$Cd$_{0.14}$Te sample~\#2  
having high electron density $n = 7.5 \times 10^{11}$ cm$^{-2}$,
see Fig.~\ref{fig_1}.
The decrease of electron density results in the lowering of the Fermi energy and, correspondingly, 
increase of the electron $g$-factor absolute value.\cite{Ivchenkobook} As a result, 
for $|\Delta_Z|> \Delta E$ only $odd$ numbers of minima can be detected. In fact, only $odd$ 
minima in a certain magnetic field range have been detected in samples \# 1, 5, and 7, which all 
have low electron density $n < 2.5 \times 10^{11}$~cm$^{-2}$.
This is seen in magnetotransport data obtained in e.g. 21~nm QWs for magnetic fields lower 
than 1.5~T as demonstrated in Fig.~\ref{fig_4}(c).

For the particular case of the linear dispersion, the cyclotron mass depends on the electron energy 
yielding nonequidistant Landau levels~\cite{CastroNeto2009,Masubuchi2013}. The SdH oscillations develop at the threshold 
field of 1.2 T which corresponds to the filling factor of 3, see Fig.~\ref{fig_4}(d) 
for sample \#3 with 6.6 nm QW (van der Pauw geometry).
%
%
With
the further field increasing a minima with filling factors 2 and 1 (not
shown) are detected. The absence of the detectable higher filling
factors in this sample is caused by the square root dependence of
the Landau level's energy being, as addressed above, characteristic for the Dirac
fermion system. Higher filling factors up to 7, both $odd$ and $even$, 
however, become visible in the carrier density dependence of the longitudinal
resistance measured at a constant magnetic field in a gated 6.6~nm QW Hall 
bar sample \#4, see Fig.~\ref{fig_6}.

\begin{table*}[tb]
\caption{
Parameters of the investigated samples at $T$ = 4.2~K. 
Second and third columns show the Hg contents in quantum well and QW width. 
The transport scattering times $\tau_{\rm tr}$ have been evaluated from the electron mobility 
and $\tau_{\rm CR}$ have been estimated from full width at half-maximum of radiation 
transmittance measurements under the cyclotron resonance conditions. 
Carrier densities given for samples \#3 and \#6 are obtained by optical 
doping. For that the structures were illuminated for time $t_{ill}$ with red LED. 
Magneto-transport measurements carried out on gated Hall 
bar sample \# 4 show that the transport relaxation time in this sample  increases with the 
rising carrier density as $\tau_{\rm tr} \propto \sqrt{n}$. This result is in full
agreement with the theory for systems characterized by the linear dispersion and 
short-range scattering, for details see Refs.~\onlinecite{Kozlov2012,dassarma11}.
}
\label{tab_1}
\begin{ruledtabular}
\begin{tabular}{c|c|c|c|c|c|c|c}
sample \# & Hg (\%) & $L_w$ (nm) & design         & $t_{ill}$ (s) & $n$ (10$^{11}$ cm$^{-2}$) & $\tau_{\rm CR}$ (ps)  & $\tau_{\rm tr}$ (ps)\\
\hline
1         & 100     & 5          & square         & -             & 2.4                        & 0.29                  & 0.24\\
\hline
2         & 86      & 8          & Hall bar       & -             & 7.5                        & -                     & 0.74 \\
\hline
\hline
3         & 100     & 6.6        & square         & 80            & 1.1                        & 0.44                  & 0.59 \\
\hline
4         & 100     & 6.6        & gated Hall bar & -             & 0.5 $\div$ 4.5               & -                     & 0.82 at $n$ = 10$^{11}$ cm$^{-2}$\\
\hline
\hline
5         & 100     & 8          & square         & -             & 2.4                        & 0.68                  & 0.65 \\
\hline
6         & 100     & 8          & square         & 80            & 3.2                        & 0.76                  & 0.68 \\
\hline
7         & 100     & 21         & square         & -             & 1.7                        & 1.4 & 1.58\\
\end{tabular}
\end{ruledtabular}
\end{table*}

\subsection{Methods}\label{subsec:methods}

For optical excitation we apply a $cw$ CH$_3$OH laser emitting a radiation with 
frequency $f\,$=$\,2.54$\,THz (wavelength $\lambda=118.8\,\mu$m) 
and $f\,$=$\,1.62$\,THz ($\lambda\,$=$\,184\,\mu$m).\cite{Kvon2008,Karch2010}
The incident radiation power $P\,$$\approx$$\,10$\,mW  is modulated at about 700~Hz by an optical chopper.
The radiation at normal incidence is focused to a spot of about 1.5\,mm diameter at the center of sample.
The spatial beam distribution has an almost Gaussian profile which is measured by a pyroelectric camera.\cite{Ziemann2000}
Photocurrent, photoresistance, and optical transmission as functions of an applied magnetic field 
$\bm B$ have been studied, applying  linearly- as well as right- ($\sigma^+$) and left- ($\sigma^-$) 
handed circularly-polarized radiation.
The corresponding experimental setups are shown in Fig.~\ref{fig_0}.
For electro-optical measurements the radiation induced electric current components $j_{x,y}$ have been 
measured via a voltage drop $U_{x, y} \propto j_{x,y}$, 
picked up across a $R_i=10$~M$\Omega$ load resistor and applying the lock-in technique.
While the photocurrent is measured in unbiased samples, to detect the photoresistance, 
we applied an external $dc$ bias voltage $U_B$ 
passing dc current of either + 0.1 or -0.1~$\mu$A.
For a pre-resistor of 10~M$\Omega$ used in these experiments this bias voltage resulted 
in the current through the sample $I_{dc} = \pm$ 0.1 $\mu$A.  
The photoresistance signal is expressed as a change of the longitudinal resistance $\Delta R_{xx}$ in the presence and in the absence of illumination.
Two methods have been used to obtain  $\Delta R_{xx}(B)$. In the first one
the photoresistive signal in response to the modulated radiation has been extracted 
from the total photovoltage making use of 
the fact that it changes the sign upon reversal of the bias polarity.
Consequently, the half of difference between the signals for 
positive and negative bias voltages 
yields the photoresistance signal, whereas the half of their  sum yields the 
bias voltage independent photocurrent strength.
In the second method, also providing the photoresistance $\Delta R_{xx}(B)$,  
we measure the longitudinal sample resistance in the dark and under the illumination with non-modulated THz radiation.

\section{Experimental Results}\label{sec:exper}

In (013)-oriented QWs excitation by normally incident THz radiation is 
known to result in photogalvanic~\cite{Wittman2010,Danilov2009} and magnetophotogalvanic 
currents.\cite{Diehl2009,Olbrich2013}
These photocurrents have already been studied in details being out of scope of the current paper. 
As addressed above here we focus on study in depth the magnetic field-induced oscillations 
of the magnetophotogalvanic current observed at low temperatures 
under the conditions of the cyclotron resonance absorption.
Note that the latter one has been widely studied in 
HgTe based materials.\cite{Meyer1992,Lovold1998,Lovold1998_2,Lovold1996,Truchsess1996,Goldmann1986,Heitmann}

\subsection{Results for parabolic dispersion}

\begin{figure}[t]
\includegraphics[width=\linewidth]{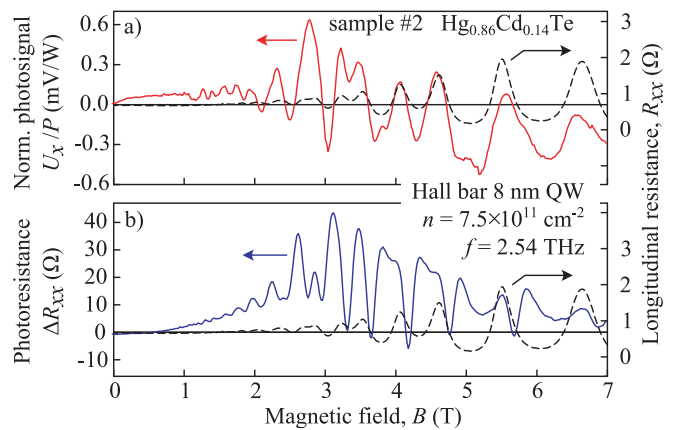}
\caption{Photosignals  obtained for an ungated 8~nm HgCdTe
Hall bar sample~\#2 
excited by linearly polarized radiation as a function of magnetic field $B$.
Panel~(a) shows the normalized by the radiation power photosignal $U_x/P$ 
induced in the unbiased sample, and panel (b) presents the photoresistance 
response $\Delta R_{xx}$ measured in the biased sample.
Black dashed curves show SdH oscillations of resistivity $R_{xx}$. 
}
\label{fig_1}
\end{figure}

\begin{figure}[h]
\includegraphics[width=\linewidth]{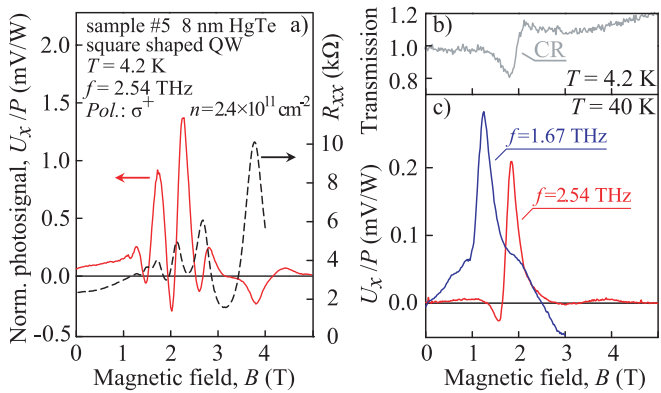}
\caption{Photosignal normalized by the radiation power, $U_x/P$, for the 
8~nm pure HgTe QW square sample \#5, excited by $\sigma^+$ radiation vs. magnetic field $B$ at (a) $T$ = 4.2 K and (c) $T$ = 40 K. 
Latter data are shown for two different frequencies, 2.54~THz (red) and 1.64~THz (blue). Panel (b) shows the change of the radiation transmission upon sweeping  magnetic field.
The data are obtained for $f = 2.54$~THz and are given in arbitrary units.
Black dashed curve in panel (a) shows SdH oscillations of resistivity $R_{xx}$.}
\label{fig_2}
\end{figure}

We start with the data obtained for the 8~nm  Hg$_{0.86}$Cd$_{0.14}$Te QWs  Hall bar sample~\#2 
and representing QWs with normal parabolic bands order.
Exciting the unbiased  sample at zero magnetic field with linearly polarized radiation we detected
the photocurrent exhibiting the characteristic behavior of the photogalvanic effect.\cite{Wittman2010,Danilov2009}
Sweeping the external magnetic field we observed that the photosignal shows
the nonmonotonous behavior superimposed with oscillations and with a maximal signal at $B \approx 2.8$~T. 
The data for $T=4.2$~K are plotted in Fig.~\ref{fig_1}(a). 
The observed oscillations correlate with $1/B$-periodic oscillations of resistivity caused by the SdH effect, 
shown by black dashed lines in panels (a) and (b). The oscillatory  behavior is also detected for the 
photoresistance signal, which, 
however, does not completely correlate with the SdH oscillations, see
Fig.~\ref{fig_1}(b). Note, that while the oscillations of photoresistance 
with similar features have been detected in HgTe-based and other low dimensional 
systems,\cite{Kropotov1987,Kvon2013_2,Pakmehr2013a,Pakmehr2013b}  the
oscillations in the photocurrents generated in unbiased samples have not been observed so far.

We attribute the observed nonmonotonic behavior of the envelope of the photosignals,
which is particularly clearly seen in the photoresistance data of Fig.~\ref{fig_1}(b), 
to the cyclotron resonance (CR).
In order to verify this conjecture we switched to a large area square-shaped samples,
which allow us, in addition to the photo-electric experiments, to measure the radiation transmission.
Studying the photocurrent and photoresistance in such samples we also observed oscillations. Figures~\ref{fig_2} -- \ref{fig_4} show the photocurrent and photoresistance responses measured in pure HgTe QW samples of different QW widths characterized by normal ($L_w$=5~nm, sample~\#1) and inverted ($L_w$=8, samples~\#5, \#6 and $L_w=21$~nm, sample~\#7) band order. These figures also present the longitudinal resistance measured in the van der Pauw geometry and the radiation transmission.
The photocurrent and photoresistance signals detected at liquid Helium temperature in sample~\#5 with 8~nm QW, which is characterized 
by the almost parabolic dispersion with inverted  band structure order, 
are shown in Figs.~\ref{fig_2}(a) and~\ref{fig_3}(b).
Like in the data for the sample~\#2 characterized by  normal band order discussed above, see Fig.~\ref{fig_1}, the photocurrent detected in sample~\#5 exhibits multiple sign inversions and is enhanced in the vicinity of the cyclotron resonance detected by the radiation transmission, see Fig.~\ref{fig_2}(b). Similar behavior is detected in the photoresistance 
$\Delta R_{xx}$, see Fig.~\ref{fig_3}(b). Photoresistance data are obtained applying two methods described in Sec.~\ref{subsec:methods}. In the first one $\Delta R_{xx}$ has been obtained applying $dc$ bias voltage of either $+1$ or $-1$~V 
and modulated radiation. In the second method  we measured longitudinal resistance $R_{xx}$ in the dark and in the presence of unmodulated THz radiation applying standard magneto-transport set-up, Fig.~\ref{fig_3}(a), and plotted the difference $\Delta R_{xx}$ in Fig.~\ref{fig_3}(c).
The data reveal that, while illumination does not change the period of the SdH oscillations it results in substantial decrease of their amplitude in the range of magnetic fields corresponding to the cyclotron resonance ($B_{\rm CR} \approx 2$~T).
Comparison of the photoresistance signal obtained by these two methods, see Fig.~\ref{fig_3}(c), shows that the results agree very well.

As the temperature increases the oscillations become less pronounced and 
almost vanish for $T =40$~K, so that both signals demonstrate a single resonance peak 
at $B_{\rm CR}= 1.8$~T for $f = 2.54$~THz and
$B_{\rm CR}= 1.2$~T for $f = 1.64$~THz. 
These data are shown for the photocurrent in Fig.~\ref{fig_2}(c).
Clear resonances at the same magnetic field strength $B_{\rm CR}$
are also detected in the transmission experiments, see Fig.~\ref{fig_2}(b) for the radiation with 
$f = 2.54$~THz. Experiments applying right-handed circularly polarized radiation revealed that a resonance dip is present for the positive magnetic fields only. Switching the radiation helicity from $\sigma^+$ to $\sigma^-$ results in the resonance for negative magnetic fields ($B_{\rm CR} = -1.8$~T at $f = 2.54$~THz, not shown). For linearly polarized radiation being the superposition of $\sigma^+$ and $\sigma^-$ photons, the resonance is observed for both magnetic field polarities.
All these facts provide the evidence that the transmission dip at $B_{\rm CR}$, as well as the photosignal increase in the vicinity of $B_{\rm CR}$, are caused by 
the cyclotron resonance. 
From the cyclotron resonance position 
\begin{equation}
\label{CR:exp}
|B_{\rm CR}| = 2\pi f \frac{m_c c}{|e|}
\end{equation}
and its full width at half maximum, both determined from the radiation transmission data, we obtain effective mass $m_c = 0.02m_0$ and the scattering time $\tau_{\rm CR} = 0.68$~ps. 
Note, that the latter value correlates well with the momentum scattering time, $\tau_{\rm tr}$,  obtained from mobility, see Table~\ref{tab_1}. Small deviations between these values detected in our experiments can be related with electron gas heating and radiative damping of the CR.\cite{mikhailov,kono}
Same results are observed for other samples with inverted (samples~\#6 and \#7) and non-inverted (sample~\#1) parabolic dispersion, see for typical curves Fig.~\ref{fig_4}(a) -- (c). It is seen that all samples show an oscillatory behavior of the photocurrent at $T$ = 4.2~K and a single peak at $T$ = 40~K, the latter corresponds to the cyclotron resonance position verified by the radiation transmission shown in the same figures.

\begin{figure}[t]
\includegraphics[width=0.8\linewidth]{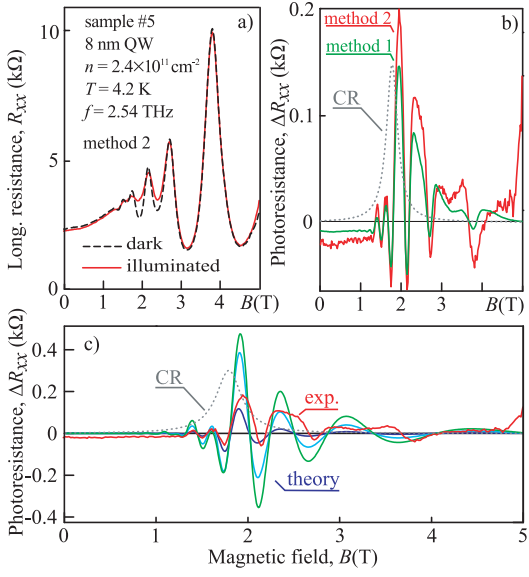}
\caption{Photoresistance and SdH data for square sample \#5 with $L_w=8$ nm. (a) Longitudinal resistance measured with (red/solid) and without (black/dotted) $cw$ THz radiation. (b) Green curve shows photoresistance, $\Delta R_{xx}$, for $dc$ bias and modulated radiation (method 1), red curve shows $\Delta R_{xx}$ for $ac$ bias and non-modulated radiation (method 2).
(c) Measured $\Delta R_{xx}$ (second method, left scale) and theoretical fit (arb. units). Dashed lines marked as ``CR'' in panels (b) and (c) show calculated cyclotron resonance absorption. The data are given in arbitrary units.
}
\label{fig_3}
\end{figure}

\begin{figure}[t]
\includegraphics[width=\linewidth]{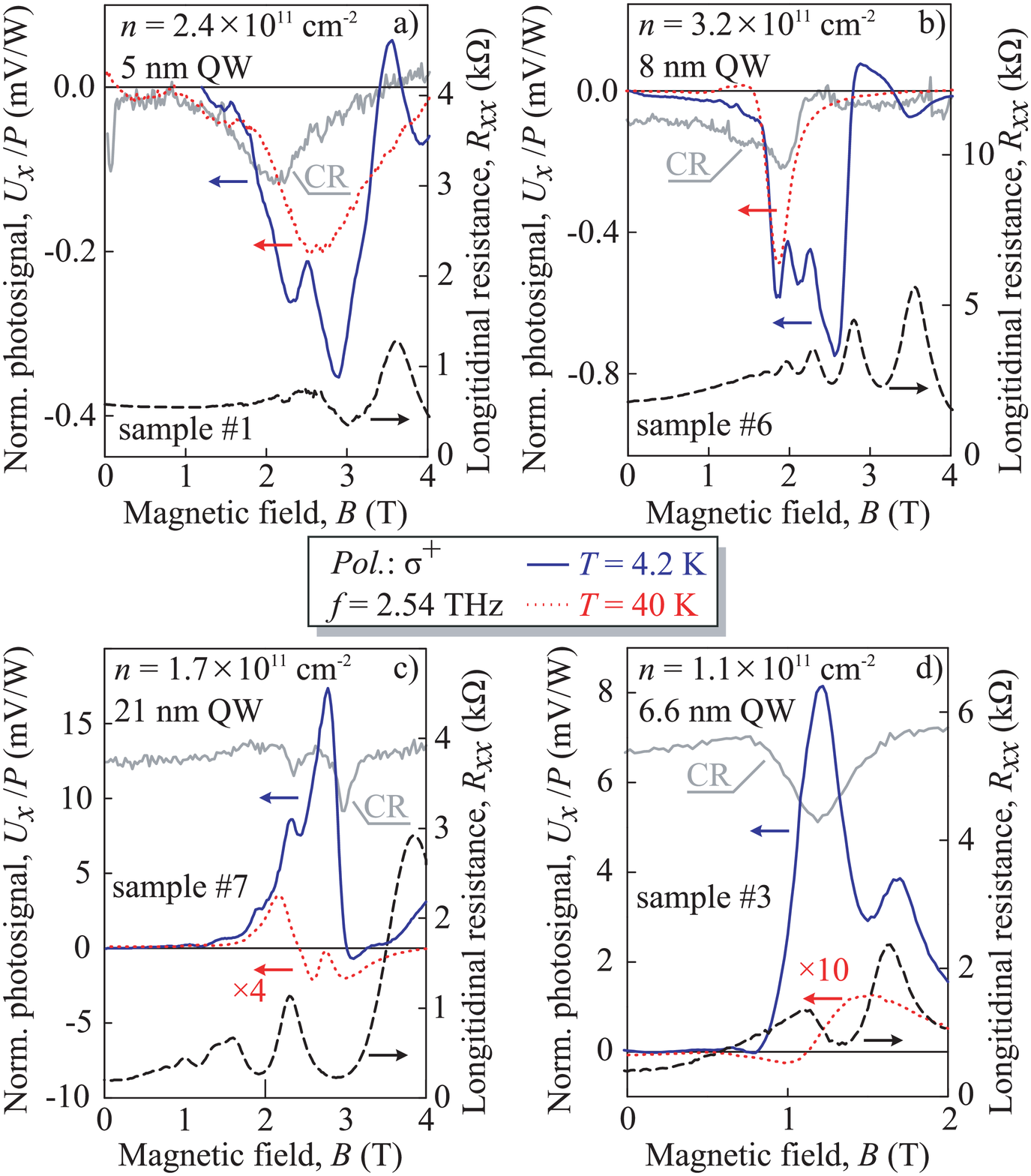}
\caption{Photosignal normalized by radiation power $U_x/P$ (left scale) and SdH oscillations (right scale) for four different QW widths, (a) $L_w$ = 5 nm, (b) $L_w$ = 8 nm, (c) $L_w$ = 21 nm, and (d) $L_w=6.6$~nm measured for two temperatures, $T = 4.2$~K (blue lines) and $T$ = 40~K (dotted red lines).
Grey lines indicated as ``CR'' show the change of the radiation transmission upon sweeping  magnetic field measured at $T=4.2$~K.
The data are given in arbitrary units.
}
\label{fig_4}
\end{figure}

\subsection{Results for a Dirac fermion system}

Now we turn to the measurements carried out 
on 6.6~nm QW samples \#3 and \#4, which are characterized by a linear energy spectra.\cite{Kvon2011,Olbrich2013}
In this system, carrier type and density have been controllable changed either by 
a gate voltage or optical illumination with red light-emitting diode~(LED) 
in ungated samples (optical doping), see Table~\ref{tab_1} where the illumination time $t_{ill}$ is indicated.
The optical doping has been obtained using the persistent photoconductivity 
effect well known for HgTe/HgCdTe QWs~\cite{Kvon2011,Ikonnikov2011,Olbrich2013}.
Figure~\ref{fig_4}(d) shows the magnetic field dependence of the photocurrent 
for sample~\#3.
Similar to the data described above and obtained for the structures with almost parabolic dispersion, see Sec. III A, 
the photocurrent measured at $T=4.2$~K  exhibits oscillations correlating with the SdH oscillations,
and the sample transmission has a clear cyclotron resonance dip at magnetic field $B = 1.2$~T.
Increasing the carrier density by the  illumination  with red LED we observed that the 
CR position, $B_{\rm CR}$, shifts to higher values by several times (not shown).
The shift of the resonance caused by energy dependence of the cyclotron mass 
for the Dirac fermion systems and variation of the Fermi energy upon the 
illumination\cite{Olbrich2013,Masubuchi2013} is described by
\begin{equation}
\label{CR:exp:lin}
|B_{\rm CR}| = \frac{(2\pi)^{3/2} \sqrt{n} \hbar c f}{|e| v_{\rm DF}}.
\end{equation}
To obtain a fine tuning of the carrier density we performed additional 
measurements on the gated samples subjected to a constant magnetic field.
The photocurrent together with the corresponding SdH oscillations detected 
in sample~\#4 is shown in Fig.~\ref{fig_6} for two values of magnetic field.
Besides the observed correlation between oscillations of the photocurrent and SdH, 
the figure indicates the non-monotonic behavior of the 
envelope function with the maximum at a density denoted as $n_{\rm CR}$. 
Performing these measurements for different values of the static magnetic field, $B$, 
we observed that
$n_{\rm CR}$ increases with rising magnetic field as 
$n_{\rm CR} \propto B^2$, see Fig.~\ref{fig_6}(c).

\begin{figure}[t]
\includegraphics[width=\linewidth]{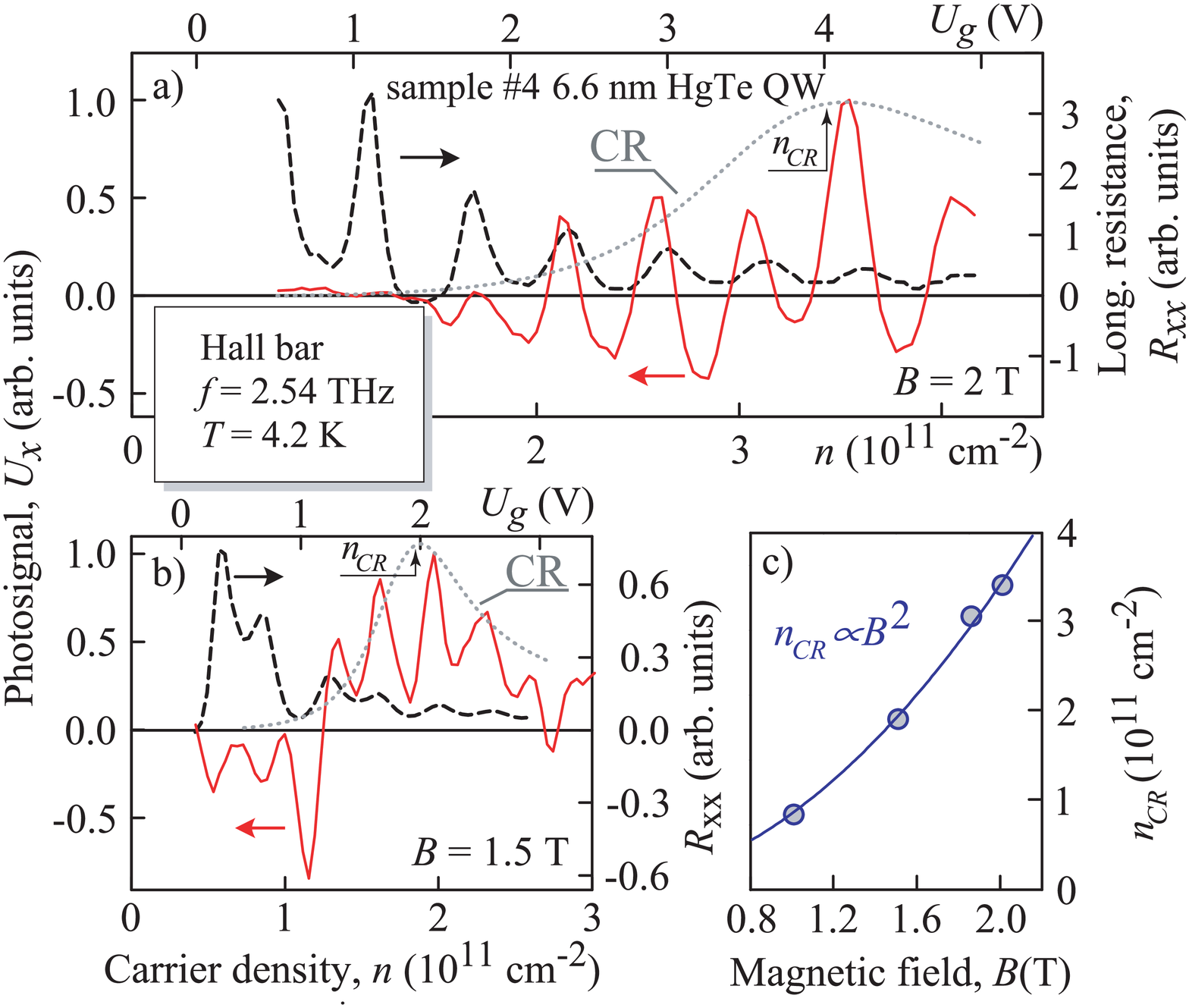}
\caption{Normalized photosignals (left scale) and longitudinal resistance $R_{xx}$
(right scale) as functions of the electron density 
(bottom scale, for the corresponding gate voltage $U_g$ see top scale) 
measured for two different magnetic field values, $B= 2$~T [panel (a)] and 1.5~T [panel (b)].
Grey lines marked as ``CR'' in panels (a) and (b) show radiation absorption 
calculated after Eq.~\eqref{sigma:ac} for two values of magnetic field
$B=2$~T and $B=1.5$~T, respectively.  Note that absorption is given in arbitrary units 
and all parameters used in the calculations are taken from experiments on magneto-transport and 
optical transmission performed for sample~\#4.
The velocity of Dirac fermions $v_{\rm DF}=7.6\times 10^{5}$~m/s for these calculations is taken
close to that experimentally determined  in Ref.~\onlinecite{Olbrich2013} and $\tau_{\rm tr} = 0.82$~ps at $n=10^{11}$~cm$^{-2}$.
(c) Magnetic field position of maximum photosignal as a function of electron density, 
showing a $n_{\rm CR} \propto B_{\rm CR}^2$ dependence. 
Solid lines in panels (a) and (b) show  radiation 
absorption calculated after Eq.~\eqref{sigma:ac}.
}
\label{fig_6}
\end{figure}

\section{Theory}\label{sec:theory}

The experiments  discussed above demonstrate that photocurrent and photoresistance
exhibit oscillations similar to with the SdH oscillations of longitudinal resistivity.
The oscillations, detected for all three types of electron dispersion,
are enhanced at the cyclotron resonance and vanish with increasing the temperature, 
showing in this case only one peak in the signal being caused by the cyclotron resonance.
In the following we present the theory describing the origin of the photogalvanic effect in classically strong magnetic fields where the electron cyclotron frequency $\omega_c$ exceeds the electron momentum scattering rate $1/\tau_{\rm tr}$, this condition is certainly fulfilled 
for $B>0.5$~T in all our samples. We show that the photocurrent oscillations, similarly to the de Haas-van Alphen and 
Shubnikov-de Haas   effects, 
stem from the consecutive crossings of Fermi level by Landau levels. The peculiarities of quantum oscillations as 
functions of magnetic field and electron density are discussed.

\subsection{Oscillations of photogalvanic current in quantizing magnetic field}

We begin with a model  of the photocurrent generation developed on the 
basis of the  spin-dependent asymmetric energy relaxation (relaxation mechanism~\cite{Ganichev2006,Ganichev2007}). In the framework of this model the 
Drude absorption of THz radiation leads to the electron gas heating.\cite{Olbrich2013} The subsequent energy relaxation 
of the heated carriers in such materials becomes spin-dependent, because the matrix element of electron 
scattering by phonons contains asymmetric spin-dependent terms.\cite{IvchenkoPikus1983,Belinicher1982,Ganichev2006} 
Figure \ref{fig:mech} sketches the cyclotron motion of electrons in the presence of asymmetric energy relaxation in 
two spin subbands $s_z=\pm 1/2$ in  the case of the classically strong magnetic field, $\omega_c \tau_{\rm tr} \gg 1$. 
We recall that
$\omega_c = |e B|/m_cc$ is the cyclotron frequency, $c$ is the speed of light, $m_c$ is the cyclotron mass given by 
$m_c = \hbar^2kdk/{dE},$  $E = E(k)$ is the electron dispersion, and $\tau_{\rm tr}$ is the momentum (transport) scattering time. 
The classical cyclotron 
orbits of electrons in the spin-up subband are shown by closed circles of the cyclotron radius $R_c = v/\omega_c$,  where 
${v \equiv v(k) = \hbar^{-1} \partial E/\partial  k}$ is the electron velocity. Without scattering the electron moves 
along the large circle shown by the solid line. We consider the dominating phonon-assisted relaxation process due to which 
the electron energy decreases. The scattering results in the displacement of the orbit center,\cite{titeica,aleiner,dmitriev:rev}  
and the orbits of smaller diameters shown by dashed and dotted lines for two selected scattering points on the orbit with positive 
and negative values components of the wavevector $k_x$.

\begin{figure}[hptb]
\includegraphics[width=0.6\linewidth]{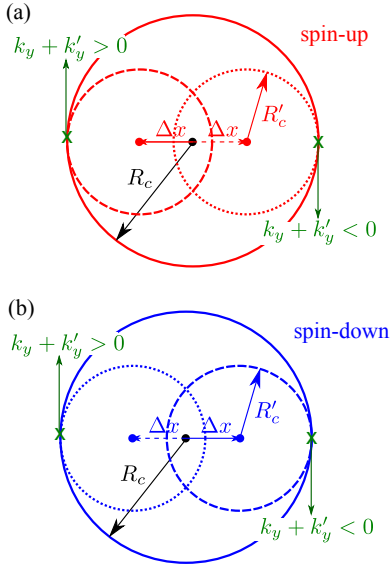}
\caption{Mechanism of the photocurrent formation in the classically strong magnetic field, $\omega_c \tau_{\rm tr} \gg 1$ in the spin-up subband (a) and in the spin-down subband (b). Solid circles depict the cyclotron orbit in the real space for the particle with given energy $E$. Cyclotron radius is $R_c$. Dashed and dotted red/blue circles depict cyclotron orbits for the electron after inelastic scattering 
by a phonon. Two possibilities for the scattering event are shown by crosses: at $k_y+k_y'>0$ and at $k_y+k_y'<0$. Due to change of the cyclotron radius, electron shifts in the real space by $\Delta x= |R_c - R_c'|$, where $R_c'$ is the cyclotron radius after the collision. The scattering processes are equally probable if wavevector-dependent contribution $\propto (k_y+k_y')$ is neglected 
in the matrix element, Eq.~\eqref{me1}. In this case any net shift of electron is absent. Allowance for the wavevector-dependent contribution in Eq.~\eqref{me1} makes scattering with $k_y+k_y'>0$ more probable giving rise to the flow in the spin-up subband directed along $x$ axis.}
\label{fig:mech}
\end{figure}

To take into account the spin-dependent scattering asymmetry we consider $\bm k$-linear terms in the electron-phonon interaction matrix elements for spins aligned along the magnetic field $\bm B$, which as we show below, are relevant for the photocurrent generation.~\cite{footnote013}
The corresponding matrix element has the form
\begin{equation}
\label{me1}
V_{\bm{k}'\bm{k}} = V_0 + V_1 
\sigma_z (k_{y} + k^\prime_{y}).
\end{equation}
where the first term in the right-hand side describes the conventional spin-independent scattering, $\sigma_{z}$ is the Pauli matrix, $\bm{k}$ and $\bm{k^\prime}$ are  the initial and final wave vectors. 
We emphasize that the above terms are allowed for gyrotropic media only~\cite{Ganichev2006,Ganichev2009DMS,GanichevBIASIA} and have been considered for HgTe-based quantum wells with both 
parabolic~\cite{Diehl2009} and linear dispersions.\cite{Olbrich2013} Evidently, the shift of orbits for positive and negative $k_{y}$ are opposite, resulting in the shifts of electrons parallel or antiparallel to ${x}$ axis. For the fixed spin, e.g. spin-up in Fig.~\ref{fig:mech}(a) spin dependent scattering makes the probabilities of these events unequal [higher for $k_y>0$ and lower for $k_y<0$, see Eq.~(\ref{me1})], which results in the steady electron flow along $x$ axis, $i_{x,+}$. For the opposite spin, see Fig.~\ref{fig:mech}(b), the situation reverses and the flow $i_{x,-}$ is oppositely directed. Consequently, in the absence of Zeeman effect we obtain a pure spin current. However, the magnetic field induced Zeeman splitting causing unequal electron subband populations and mobilities in each spin subband makes the magnitudes of the flows unequal giving rise to  the net \emph{dc} current, $j_x=e(i_{x,+} + i_{x,-})$. 

Quantum effects enter the picture as an interference of electron waves on classical orbits yielding Landau levels. As it is well known, the quantization results in $1/B$ oscillations of density of states and of scattering rates caused by the crossing of the Fermi level, $E_F$, by Landau levels.\cite{aleiner,dmitriev:rev,abrikosov}
In particular, we show that oscillations of the photocurrent stem from periodic variation of the radiation absorption rates, occupations of spin-up and spin-down subbands, $n_\pm$, and electron scattering rates $W^\pm_{\bm k'\bm k}$.

Formally, the electron fluxes in $x$-direction are given by the product of elementary displacement of the charge carrier in the real space due to the scattering event and its probability
 \begin{equation}
\label{displacement}
	i_{x,\pm} = \sum_{\bm k, \bm k'} (x_{\bm k} - x_{\bm k'}) W^\pm_{\bm k' \bm k},
\end{equation}
where the position of the cyclotron orbit center is given by
\begin{equation}
\label{center}
	x_{\bm k} = \hbar k_x {c \over |eB|}.
\end{equation}
These expressions are general and valid for parabolic as well as for linear dispersions.\cite{Schliemann2008} 
Considering the scattering on phonons and assuming that at low temperatures the thermalization between the spin branches due to electron-electron collisions is not efficient\cite{footnote_ee} we obtain the following expression for the total \emph{dc} current 
\begin{equation}
\label{j:tot}
j_x =e(i_{x,+} + i_{x,-}) = {e\beta \zeta c \over |B|} |E_0|^2 [n_+\mu_+(\omega) - n_- \mu_-(\omega) ],
\end{equation}
where $E_0$ is the complex amplitude of the incident radiation, $\beta =1$ for parabolic 
and $\beta=1/2$ for linear energy dispersion, $\mu_\pm(\omega)$ are the high-frequency 
electron mobilities in each spin branch related with the high-frequency (\emph{ac}) dissipative conductivities as $\sigma_\pm(\omega) = |e|n_\pm\mu_{\pm}(\omega)$, and 
 $$\zeta = \frac{2{\rm Re}(V_0 V_{1}^*)}{|V_0|^2} \frac{k}{v(k)}$$ 
is the small parameter responsible for the scattering asymmetry.~\cite{footnote_xi}
It follows from Eq.~\eqref{j:tot} that the photocurrent is proportional to the squared amplitude of the electromagnetic field, i.e. to the radiation intensity, as well as to the radiation absorption rate in the corresponding spin subband $\propto  n_\pm \mu_\pm(\omega)$.
We stress that Eqs.~\eqref{displacement},~\eqref{center}, and \eqref{j:tot} hold in both classical and quantizing magnetic fields provided that $\hbar\omega_c \ll E_F$, where $E_F$ is the Fermi energy, if $\bm k$, $\bm k'$ are replaced by the appropriate quantum numbers in the magnetic field, namely, $N$ (the Landau level number) and $P_x = \hbar k_x - e A_x/c$ the generalized momentum ($A_x =- By$ is the vector potential of the static field), and all quantities are expressed via $N$, $N'$, $P_x$, $P_x'$.\cite{Lifshitz1981}  The quantum oscillations of the photocurrent $\bm j$, Eq.~\eqref{j:tot}, originate from the oscillations of $n_\pm$ and $\mu_\pm$. 

It is convenient to decompose the current $j$ as a sum of two components:
one, $j_n$, related to the spin polarization $S_z = (n_+ - n_-)/2n$ in the system and the other, $j_\mu$, related solely with the difference, $\mu_+(\omega) - \mu_-(\omega)$, of high-frequency mobilities in spin subbands; total electron density $n = n_+ + n_-$. To do this we
rewrite the term in square brackets in Eq.~\eqref{j:tot} in form 
$$S_z n[\mu_+(\omega) + \mu_-(\omega)] + n\frac{\mu_+(\omega) - \mu_-(\omega)}{2}.$$
Then
\begin{equation}
\label{j:tot1}
j_x = j_n + j_\mu,
\end{equation}
where 
\begin{equation}
\label{j1}
j_n = {2e \zeta c \beta |E_0|^2 \over |B|}
S_z n \mu(\omega),
\end{equation}
\begin{equation}
\label{j2}
j_\mu = {e \zeta c \beta |E_0|^2 \over |B|}
 n\frac{\mu_+(\omega) - \mu_-(\omega)}{2},
\end{equation}
and $\mu(\omega)=[\mu_+(\omega) + \mu_-(\omega)]/2$. 
The above equations show that the oscillatory part of the photocurrent 
is, in fact, contained in the magnetic field dependence of (i) the electron spin polarization, $S_z$ (contribution $j_n$), and (ii)
the mobility difference in subbands with opposite spins, $\mu_+(\omega) - \mu_-(\omega)$ (contribution $j_\mu$). Thus, below we focus on these quantities and derive the expressions for them in the cases of QWs with parabolic and linear energy dispersions.

\subsection{Photocurrent in the system with parabolic dispersion}\label{sec:parab}

To be specific we first analyze the case of parabolic dispersion characterized by the energy-independent effective mass $m$, and, in line with experiment, the frequency range of $\omega$ in the vicinity of the cyclotron frequency $\omega_c$. 

We start with calculation of electron spin $S_z$ determining the first contribution to 
the photocurrent, $j_n$. With allowance for the quantum oscillations (de~Haas-van~Alphen effect), it is given by~[cf. Ref.~\onlinecite{bychkov60}]
\begin{equation}
\label{Sz0}
S_z = \frac{1}{2} \frac{\int_0^{E_F} [\nu_+(E) - \nu_-(E)]  dE}{\int_0^{E_F} [\nu_+(E) + \nu_-(E)]  dE},
\end{equation}
where $\nu_\pm(E) = \nu_0(E \mp \Delta_Z/2)$ are the densities of states in each spin branch with $\nu_0(E)$ being the density of states per spin found neglecting Zeeman effect.  Here $\Delta_Z = g\mu_B B$ is the Zeeman splitting with $g$ being the electron Land\'{e} factor and $\mu_B$ being Bohr magneton. 
We consider classical magnetic fields, where $\hbar\omega_c \ll E_F$ and for $E_F\tau_{q}/\hbar \gg 1$.
Here $\tau_q$ is the quantum scattering time related to Dingle temperature, 
which describes the lifetime of an electron in a given quantum state, and is shorter than the transport time $\tau_{\rm tr}$ for a smooth disorder potential. 
Under these assumptions the density of states can be written as\cite{dmitriev:rev,ando74,willander} 
\begin{multline}
\label{DOS}
\nu_0(E) = \frac{m}{2\pi\hbar^2}   \left[ 1- 2\exp{\left(- \frac{\pi}{\omega_c \tau_q}\right)}
\cos{\left( 2\pi \frac{E}{\hbar\omega_c} \right)} \right],
\end{multline}
where we have taken into account the oscillating contributions of the first in small parameter $\exp{(-\pi/\omega_c\tau_q)}$ describing the interference of electron waves on the cyclotron orbits.\cite{aleiner} The spin polarization is proportional to the energy-integrated difference of densities of states in spin-up and spin-down subbands, Eq.~\eqref{Sz0}, namely, 
\begin{multline}
\label{Sz}
	S_z=-\frac{1}{4E_F} 
	\biggl[\Delta_Z  -\\
	 {2\hbar \omega_c \over \pi} \mathcal T_1 \sin\left({\pi \Delta_Z \over \hbar \omega_c}\right)\cos\left({2\pi E_F \over \hbar \omega_c}\right) \mathrm e^{-{\pi/ \omega_c \tau_q}} \biggr].
\end{multline}
Note that, hereinafter we neglect the oscillatory corrections of higher orders in small parameters $\hbar\omega_c/E_F\ll 1$ and $\Delta_Z/E_F\ll 1$. The factor 
\begin{equation}
\label{Tk}
\mathcal T_1 =  \frac{2\pi^2 k_B T_e}{\hbar\omega_c \sinh{(2\pi^2 k_B T_e/\hbar\omega_c)}}, 
\end{equation}
takes into account a thermal spread of the electron distribution function with $T_e$ being the electron gas temperature. Interestingly, the oscillatory contribution to the spin polarization $S_z$ being proportional to the factor $\sin{(\pi \Delta_Z/\hbar\omega_c)}$,  is absent if Zeeman splitting, $\Delta_Z$, is a multiple of the inter-Landau level distance, $\hbar\omega_c$. In this case the Landau levels in both spin branches are aligned\cite{tarasenkoZ} and oscillations of spin polarization vanish.

Now we turn to the high-frequency mobilities in each spin subband whose difference gives rise to $j_\mu$. In line with Refs.~\onlinecite{dmitriev:rev,dmitriev:cr,raichev2008} we have for $\omega$ in the vicinity of $\omega_c$
\begin{multline}
\label{mu:ac}
\mu_{\pm}(\omega) = \frac{|e|\tau_{\rm tr}/2m}{1+(\omega-\omega_c)^2\tau_{\rm tr}^2}\times \\
\left[1+ \mathcal T_1 \mathrm e^{-\pi/\omega_c\tau_q} \cos{\left( \frac{2\pi E_{F_{\pm}}}{\hbar\omega_c} \right)} F(\omega\tau_{\rm tr},\omega_c\tau_{\rm tr}) \right],
\end{multline}
with $E_{F_\pm} =  E_F \mp \Delta_Z/2$ and 
\begin{multline}
\label{f1}
F(\omega\tau_{\rm tr},\omega_c\tau_{\rm tr}) = \frac{2(\omega-\omega_c)^2\tau_{\rm tr}^2}{1+(\omega-\omega_c)^2\tau_{\rm tr}^2}\frac{\sin{2\pi\omega/\omega_c}}{2\pi\omega/\omega_c}+ \\
\frac{1+3(\omega-\omega_c)^2\tau_{\rm tr}^2}{1+(\omega-\omega_c)^2\tau_{\rm tr}^2}\frac{\sin^2{\pi\omega/\omega_c}}{(\omega-\omega_c)\tau_{\rm tr}\pi\omega/\omega_c}
\end{multline}
is the smooth (on the scale of $\hbar/E_F$) function of $\omega\tau_{\rm tr}$ and $\omega_c\tau_{\rm tr}$. Here we neglect a background contribution to the mobility resonant at $\omega = - \omega_c$.\cite{footnote_super}
Equation~\eqref{mu:ac} shows that the high-frequency mobilities contain both the 
classical (Drude) CR part, being proportional to $\tau_{\rm tr}/[1+(\omega - \omega_c)^2\tau_{\rm tr}^2]$, 
and $1/B$-periodic oscillatory contributions resulting from the consecutive crossing of Fermi level by Landau levels.\cite{dmitriev:cr,dmitriev:rev,ando75}

Taking into account Eq.~\eqref{Sz0} and Eq.~\eqref{mu:ac} 
we can obtain from Eqs.~\eqref{j1}  and \eqref{j2} 
contributions $j_n$ and $j_\mu$ in the form:
\begin{multline}
\label{j:1st}
j_n = - {e \zeta c \beta |E_0|^2 n \mu(\omega) \over 2E_F|B|}  
	\biggl[\Delta_Z  -\\
	 {2\hbar \omega_c \over \pi} \mathcal T_1 \sin\left({\pi \Delta_Z \over \hbar \omega_c}\right)\cos\left({2\pi E_F \over \hbar \omega_c}\right) \mathrm e^{-{\pi/ \omega_c \tau_q}} \biggr], 
\end{multline}
and
\begin{multline}
\label{j:2nd}
j_\mu = {e^3 \zeta c |E_0|^2 n \over 2m|eB|} \frac{\tau_{\rm tr}}{1+(\omega-\omega_c)^2\tau_{\rm tr}^2} \times \\ 
\mathcal T_1 \mathrm e^{-\pi/\omega_c\tau_q} \sin{\left( \frac{2\pi E_{F}}{\hbar\omega_c} \right)} \sin{\left( \frac{\pi \Delta_Z}{\hbar\omega_c} \right)} F(\omega\tau_{\rm tr},\omega_c\tau_{\rm tr}) ,
\end{multline}
respectively. Equations~\eqref{j:1st} and \eqref{j:2nd} describe the classical smooth part and quantum oscillations of the photocurrent in systems with parabolic dispersion.

\subsection{Photocurrent in the system with linear dispersion}\label{sec:linear}

In the HgTe-based quantum wells of critical thickness the electron energy 
spectrum at zero magnetic field is linear, $E(k) = \hbar v_{\rm DF} k$. 
Our theoretical treatment demonstrates that the basic mechanism of the photocurrent 
generation is the same as in the HgTe-based QWs with normal or inverted parabolic band structure.
The photocurrent contains two contributions $j_n$  and $j_\mu$ resulting from the different populations of spin subbands and from difference in high-frequency mobilities in these subbands and described by Eqs.~\eqref{j1} and \eqref{j2}, respectively. In the case of linear dispersion, however, the Landau levels are not equidistant and the quantum oscillation pattern changes. Here we present the set of formulae which generalize Eqs.~\eqref{DOS} -- \eqref{mu:ac} to the case of linear dispersion taking into account, as in Sec.~\ref{sec:parab}, only first order oscillatory contributions. The density of states assumes the form\cite{briskot} (see also Ref.~\onlinecite{sharapov})
\begin{multline}
\label{DOS:D}
\nu_0(E) = \frac{m_c}{2\pi\hbar^2}   \left[ 1 +  2\exp{\left(- \frac{\pi}{\omega_c \tau_q}\right)}
\cos{\left( \frac{\pi  E}{\hbar\omega_c} \right)} \right]
\end{multline}
where both cyclotron mass, $m_c \equiv m_c(E) = E/v_{\rm DF}^{2}$ and cyclotron frequency, $\omega_c \equiv \omega_c(E)= eB v_{\rm DF}^2/(E c)$, are the functions of electron energy. Under the same approximations as for the parabolic spectrum we obtain the electron spin polarization in the first order in $\exp{(-\pi/\omega_c\tau_q)}$:
\begin{multline}
\label{Sz:D}
	S_z=-\frac{1}{2E_F} 
	\biggl[\Delta_Z  +\\
	 {2\hbar \omega_c \over \pi} \mathcal T_1 \sin\left({\pi \Delta_Z \over \hbar \omega_c}\right)\cos\left({\pi E_F \over \hbar \omega_c}\right) \mathrm e^{-{\pi/ \omega_c \tau_q}} \biggr],
\end{multline}
and the high-frequency mobilities in spin-up and spin-down branches
\begin{multline}
\label{mu:ac:D}
\mu_{\pm}(\omega) = \frac{e\tau_{\rm tr}/2m_c}{1+(\omega-\omega_c)^2\tau_{\rm tr}^2}\times \\
\left[1- \mathcal T_1 \mathrm e^{-\pi/\omega_c\tau_q} \cos{\left( \frac{\pi E_{F_{\pm}}}{\hbar\omega_c} \right)} F(\omega\tau_{\rm tr},\omega_c\tau_{\rm tr}) 
\right],
\end{multline}
where function $F(\omega\tau_{\rm tr},\omega_c\tau_{\rm tr})$ is given by Eq.~\eqref{f1}.
\cite{misprint}
In Eqs.~\eqref{Sz:D} and \eqref{mu:ac:D} one has to put $E=E_F$ in $m_c$, $\omega_c$, $\tau_q$, and $\tau_{\rm tr}$.
Note that in the case of linear dispersion even short-range scattering results in different quantum (out-scattering) time $\tau_q$ and transport time $\tau_{\rm tr}=2\tau_q$. Equations~\eqref{Sz:D} and \eqref{mu:ac:D} allow us to calculate the photocurrent contributions $j_n$ and $j_\mu$ by Eqs.~\eqref{j1} and \eqref{j2} for the system with linear energy dispersion.

\section{Discussion}

The theory discussed in the previous Section allows us to describe the
experimental data. First we discuss the basic features of the photocurrent in the quantum wells with parabolic dispersion, then we address peculiarities of the quantum wells of critical thickness, particularly, the appearance of the cyclotron resonance in the photocurrent as a function of the carrier density. The photoresistivity effect is discussed in the end of this section as well.

\subsection{Quantum oscillations of photocurrent in QWs with parabolic dispersion}

\begin{figure}[hptb]
\includegraphics[width=\linewidth]{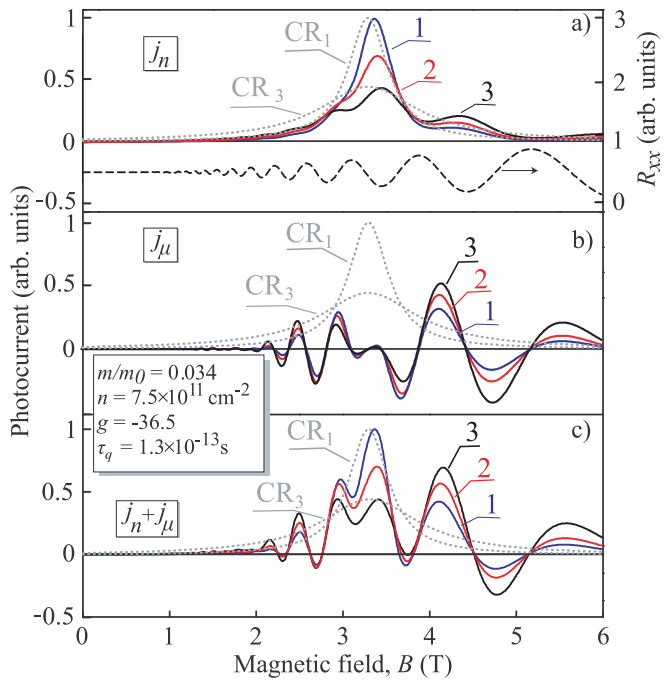}
\caption{Panels (a) and (b)
show  photocurrent contributions $j_n$ and $j_\mu$ calculated
after Eqs.~\eqref{j:1st} and \eqref{j:2nd}, respectively. The 
total photocurrent given by the sum of $j_n$ and $j_\mu$, see Eq.~\eqref{j:tot1},
is presented in panel (c).
The parameters used in the calculations are chosen close to those of 
sample~\#2:  $m=0.034m_0$, $n=7.5\times 10^{11}$~cm$^{-2}$, $g=-36.5$, 
and $\tau_q=1.3\times 10^{-13}$~sec. Curves 1, 2 and 3 are calculated for
$\tau_{\rm tr}=3\times 10^{-13}$,   $5\times 10^{-13}$ and $7.4 \times 10^{-13}$~sec, respectively.  
The normalization is the same for all panels. 
Dashed lines  show calculated cyclotron resonance absorption profiles CR$_{\rm 1}$ and  CR$_{\rm 3}$ 
obtained for $\tau_{\rm tr}=3\times 10^{-13}$   and $7.4 \times 10^{-13}$~sec, respectively. 
These data are given in arbitrary units with the CR maximum 
almost equal to  $j_n$ maximum.
Dashed line in (a) shows the first harmonic in the oscillatory 
part of $R_{xx}$. 
}
\label{fig_8}
\end{figure}

To analyze the oscillations of the photocurrent generated in the structures with almost parabolic dispersion
we calculated the individual contributions $j_n$ and $j_\mu$, 
as well as the total photocurrent $j = j_n + j_\mu$ using Eqs.~\eqref{j:1st}, \eqref{j:2nd}, and Eq.~\eqref{j:tot1}, respectively.
Figure~\ref{fig_8} shows the results obtained using an effective mass $m = 0.034 m_0$, carrier density $n = 7.5 \times 10^{11}$ cm$^{-2}$, and the electron $g$-factor $g=-36.5$, i.e. the parameters close to that of sample \#2 with 8~nm QW. Note, that large values of electron $g$-factors in HgTe-based QWs were reported in Ref.~\onlinecite{Zhang2004}.
Comparison of the total photocurrent  plotted in Fig.~\ref{fig_8} (c) 
with experimental result for sample~\#2 representing 
QWs with the normal band order [Fig.~\ref{fig_1}(a)] 
clearly shows that the main experimental features are fully reproduced 
by the theoretical calculations.
First of all, both calculated and experimental signals show pronounced 
oscillations accompanied with inversion of the current direction. They start 
to be observable in the range of magnetic fields where the cyclotron resonance takes place [see dashed line in Fig.~\ref{fig_8}(a)]. Moreover, both experiment and theory reveal that the oscillations for magnetic fields $B>|B_{\rm CR}|$ are substantial being comparable with that in the vicinity of the cyclotron resonance. Different magnetic field
behavior of the photocurrent caused by spin polarization,
$j_n$, and that driven by the difference of mobilities 
in the spin branches,  $j_\mu$, provides a way to distinguish 
their contributions to the total photocurrent. 
Figures~\ref{fig_8}(a) and (b) show that,
while $j_n$ almost follows the cyclotron resonance, being slightly modulated by the periodic sign-conserved oscillations, $j_\mu$, in contrast, 
is characterized by the multiple reversal of the current direction with a maximum far away from CR. Moreover, the latter contribution vanishes at CR position.
Figure~\ref{fig_TE_parab} shows that while $j_n$ contribution to 
the photocurrent achieves maximum at CR position and rapidly decreases 
outside the resonance, the oscillations of this term are almost absent. 
By contrast, the contribution $j_\mu$ 
vanishes exactly in the resonance and shows oscillations, which remain
even for magnetic fields beyond the resonance. 
 Experimental data shown in Fig.~\ref{fig_1}(a) reveal that $j_n$ and $j_\mu$ components of the photocurrent are comparable. 
In particular, the multiple sign inversion of the photocurrent, being 
the fingerprint of $j_\mu$ is clearly detected demonstrating that this
mechanism dominates in the total current. 
Same results are obtained for sample~\#5 representing structures with 
the inverted band order, see Fig.~\ref{fig_2}(a), being in agreement with 
the corresponding calculations, see Fig.~\ref{fig_TE_parab}.

\begin{figure}[tb]
\includegraphics[width=\linewidth]{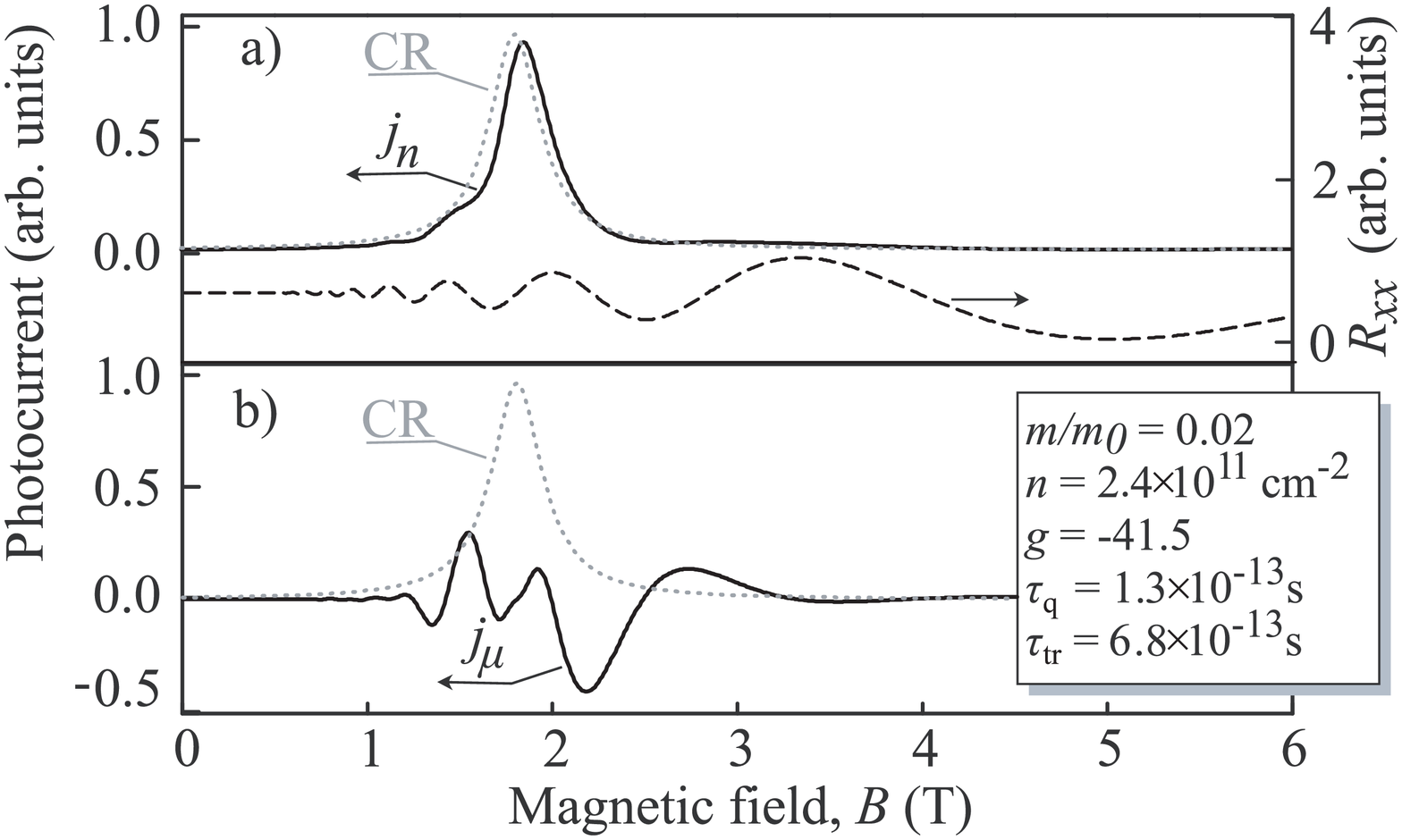}
\caption{
Panels (a) and (b)
show  photocurrent contributions $j_n$ and $j_\mu$ calculated
after Eqs.~\eqref{j:1st} and \eqref{j:2nd}, respectively.
Dotted lines show calculated cyclotron resonance absorption.
Photocurrent contributions and CR profile are normalized to 
their maximum values.
Dashed line in (a) shows first order oscillatory 
contribution to $R_{xx}$. 
The parameters used in the calculations are chosen close to those of 
sample~\#5: $m=0.02m_0$, $n=2.4\times 10^{11}$~cm$^{-2}$, $g=-41.5$, 
$\tau_q=1.3\times 10^{-13}$~sec, $\tau_{\rm tr}=6.8\times 10^{-13}$~sec.
}
\label{fig_TE_parab}
\end{figure}

\begin{figure}[tb]
\includegraphics[width=\linewidth]{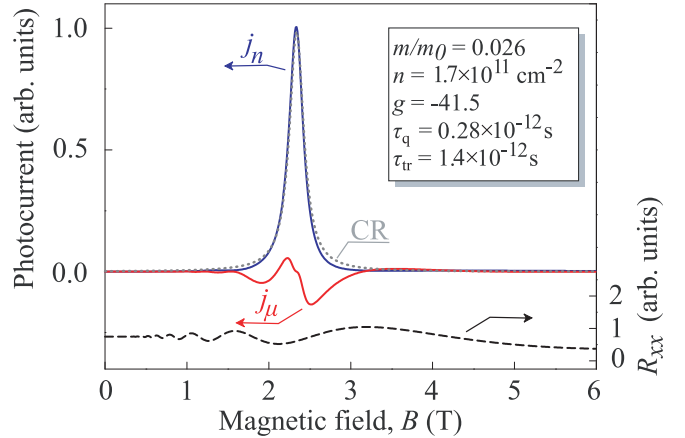}
\caption{
Photocurrent contributions $j_n$ and $j_\mu$ calculated
after Eqs.~\eqref{j:1st} and \eqref{j:2nd}, respectively.
Dotted line  shows calculated cyclotron resonance absorption.
Absorption and $j_n$ contribution to the photocurrent are normalized to 
their maximum values.
Dashed line shows first order oscillatory 
contribution to $R_{xx}$. 
The parameters used in the calculations are chosen close to those of 
sample~\#7: $m=0.026m_0$, $n=1.7\times 10^{11}$~cm$^{-2}$, $g=-41.5$, 
$\tau_q=0.28\times 10^{-12}$~sec, $\tau_{\rm tr}=1.4\times 10^{-12}$~sec.
}
\label{fig_TE_narrow}
\end{figure}

While Figs.~\ref{fig_8} and~\ref{fig_TE_parab}  present the calculations for a relatively broad cyclotron resonance,  
Fig.~\ref{fig_TE_narrow} shows the results of calculation for relatively 
narrow CR which covers only one period of SdH oscillations, see
dotted and dashed lines in Fig.~\ref{fig_TE_narrow} for CR and SdH oscillations, respectively. 
This condition is relevant to the low temperature measurements 
in QWs with normal ($L_w = 5$~nm) and inverted ($L_w = 8$ and 21~nm) parabolic bands 
summarized in Fig.~\ref{fig_4}(a)-(c). The parameters used in the calculations are chosen close to those of 
sample~\#7. The results of the calculations describe well the experimental
findings, see  Fig.~\ref{fig_4}(c).
Comparison of the calculated   $j_n$ and $j_\mu$ 
with experimental data shown in Fig.~\ref{fig_4}(a)-(c) 
demonstrates that while both contributions can be clearly identified 
the main input in these structures comes from $j_n$. 
This conclusion is proved by the higher signals at CR position 
as well as by a small number of the detected  
oscillations.~\cite{footnote_Dyakonov}

Below we analyze in more detail a complex picture of the photocurrent oscillations.
We start with the photocurrent $j_n$, which is proportional to the 
electron spin polarization $S_z$, given by Eq.~\eqref{j:1st} and 
plotted  in Figs.~\ref{fig_8} and~\ref{fig_TE_parab}, panels (a), see also blue solid curve in Fig.~\ref{fig_TE_narrow}. This contribution to the photocurrent 
represents the \emph{dc} current generated due to a difference of spin-up and spin-down subbands 
populations in the presence of magnetic field, see Eq.~\eqref{j1}. The effect is also known as a zero-bias spin separation\cite{Ganichev2006} converted to the electric current due to Zeeman effect, which has been observed in many semiconductor low-dimensional systems, for review see~Refs.~\onlinecite{GanichevBIASIA,Belkov2008,Belkovbook}. One can see that the 
dominating input comes from a 
smooth non-oscillatory part which almost follows the cyclotron resonance and is 
in agreement with Eqs.~(13) and~(14) of 
Ref.~\onlinecite{Olbrich2013}. However, since both $S_z$ and $\mu(\omega)$ in 
Eq.~\eqref{j1} contain also $1/B$-periodic components, 
the $j_n$ contains relatively small [for $\exp{(-\pi/\omega_c\tau_q)} \ll 1$] 
quantum oscillations described by $\cos{(2\pi  E_F/\hbar\omega_c)}$ superimposed 
over smooth background, as demonstrated in Fig.~\ref{fig_8}(a).
Similarly to 
de Haas-van Alphen and 
Shubnikov-de Haas effects scattering processes and thermal spread of electron 
distribution function suppress the oscillations of $S_z$ and $\mu_\pm$, see Eqs.~\eqref{Sz} and~\eqref{mu:ac}. Therefore, only smooth part of $j_n$ contribution remains responsible for the observed resonant photocurrent at high temperatures $T\simeq 40$~K, where oscillations vanish, see Fig.~\ref{fig_2}(b) and Fig.~\ref{fig_4}.

Now we turn to the second contribution, $j_\mu$, which results from the magnetic field induced difference 
of high-frequency mobilities in spin-up and spin-down branches, see Eqs.~\eqref{j2} and \eqref{mu:ac}.
It is given by Eq.~\eqref{j:2nd} and plotted in Figs.~\ref{fig_8} and Fig.~\ref{fig_TE_parab}(b) as well as red solid curve in Fig.~\ref{fig_TE_narrow}.
As seen from the figure this contribution also oscillates as a function of magnetic field, 
but compared to $j_n$ does not have any smooth part and demonstrates multiple sign inversions. 
The latter is due to the fact that the difference of mobilities $\mu_\pm{(\omega)}$ mainly comes 
from oscillatory factors $\sin{(2\pi  E_F/\hbar\omega_c)}$.\cite{footnote_smooth} In other words, 
the direction of $j_\mu$ current is determined by the electron flux in the spin branch with 
extremal density of states at the Fermi level. As magnetic field changes, either spin-up or 
spin-down branch dominates, and the photocurrent changes its direction. Moreover, $j_\mu$ 
cancels at exact resonance position, $B=B_{\rm CR}$, and the oscillation amplitude substantially 
increases for higher magnetic fields. The latter is due to an increase of quantum parameter $\exp{(-\pi/\omega_c\tau_q)}$, which governs the amplitude of the magneto-oscillations in Eqs.~\eqref{j:1st} and \eqref{j:2nd}, resulting also in the increase of Shubnikov-de Haas oscillations amplitude in the resistivity of the QW structure.  

Theoretical Fig.~\ref{fig_8} as well as experimental Figs.~\ref{fig_1}(a), \ref{fig_2}(a) show that $j_n$ and $j_\mu$ 
contributions have at low temperatures close magnitudes. It stems from the fact that $j_n$ contains small parameter $\Delta_Z/E_F$ since this term is proportional to the spin polarization $S_z$. By contrast, $j_\mu$ contains the small quantum parameter, namely, $\mathrm{exp}({-\pi/\omega_c\tau_q})$, 
while Zeeman splitting $\Delta_Z$ enters only as a ratio $\Delta_Z/\hbar\omega_c$, which can be on the 
order of unity. For typical  parameters $\Delta_Z/\hbar\omega_c\sim 1$, 
$\mathrm{exp}({-\pi/\omega_c\tau_q}) = 0.1\ldots 0.5$ the contributions are comparable.

The total photocurrent is given by the sum of $j_n$ and $j_\mu$ contributions and demonstrates rather complex oscillatory behavior shown in 
Fig.~\ref{fig_8}(c). Particularly, the oscillations 
are accompanied with sign inversion, but they are superimposed over smooth CR-like background. It is worth to stress that in addition to the 
considered above photocurrent $j = j_n + j_\mu$ 
resulting from the spin-dependent asymmetric energy relaxation mechanism the measured photocurrent may contain 
contributions caused by the spin dependent asymmetry of optical 
transitions (excitation mechanism, see Ref.~\onlinecite{Ganichev2006}). The latter mechanism
will also results in the magneto-oscillations, and the functional form of 
its individual contributions is to that of $j_n$ and $j_\mu$.
Hence, for detailed comparison of experimental data and theory one needs to take into account these additional photocurrent contributions as well as higher order terms in Eqs.~\eqref{j:1st} and \eqref{j:2nd} in quantum parameter $\exp{(-\pi/\omega_c\tau_q)}$, as experimental data presented in Figs.~\ref{fig_1} and \ref{fig_3}  reveal second harmonic in Shubnikov-de Haas oscillations of resistivity see also Sec.~\ref{sec:photocond}.

%

\begin{figure}[hb]
\includegraphics[width=\linewidth]{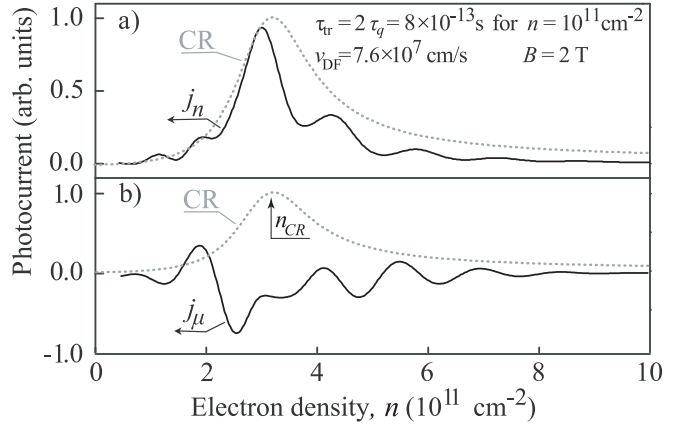}
\caption{
Panels (a) and (b) show  photocurrent 
contributions $j_n$ and $j_\mu$ calculated
after Eqs.~\eqref{j1} and \eqref{j2}, respectively. 
Spin polarization and the high frequency mobilities 
in the opposite spin branches are calculated  for $B=2$~T
after~\eqref{Sz:D} and \eqref{mu:ac:D} applying
velocity of Dirac fermions $v_{\rm DF}=7.6\times 10^{7}$~cm/s close to that
experimentally determined  in Ref.~\onlinecite{Olbrich2013}.
In the calculations we assume the short-range scattering 
with $\tau_q= \tau_{\rm tr}/2 =4\times 10^{-13}$~sec for electron density $n=10^{11}$~cm$^{-2}$.
Dotted lines show CR  profile as 
a function of electron density. The radiation absorption 
is calculated after Eq.~\eqref{sigma:ac} for 
$B=2$~T. All parameters used in the calculations are taken from 
experiments on magneto-transport and  optical transmission performed for the sample~\#4.
Note that the photocurrent contributions and CR profile are normalized to 
their maximum values.
}
\label{fig_9}
\end{figure}

\subsection{Quantum oscillations of photocurrent in QWs with linear dispersion}

Now we turn to the system with Dirac dispersion realized in quantum 
wells of critical thickness, $L_w=6.6$~nm (samples \# 3 and 4). 
The analysis performed in Sec.~\ref{sec:linear} shows that 
the magnetic field dependence of the photocurrent is 
similar to that in the system with normal or inverted parabolic band. 
This result is in agreement with the experimental data presented in Fig.~\ref{fig_4}(d). 
Comparison of Eqs.~\eqref{DOS:D} and \eqref{DOS} for photocurrent in the systems
with linear and parabolic dispersion, respectively, shows that the only differences 
between them are the amplitude and the phase of quantum oscillations.
The former one is primary determined by strong dependence of 
the cyclotron mass on the electron energy in QWs of critical thickness
which is almost absent in systems with parabolic dispersion. 
As to the phase of quantum oscillations, it is shifted by $\pi$
for Dirac fermion systems compared to that in parabolic band.

A distinguishing feature of the photocurrent in the Dirac fermions system, however, 
is the fact that the CR-resonance related effects change
upon the 
carrier density variation. It becomes possible due to the linear 
energy spectrum resulting in the dependence of the cyclotron frequency 
on the electron Fermi energy $E_{\rm F} \propto \sqrt{n}$.
The  CR-resonance as a function of the carrier density or gate voltage
can be understood by considering the electron density dependence 
of the high-frequency conductivity. The latter determines the radiation 
absorption and, in the vicinity of the cyclotron resonance, 
can be recast as
\begin{equation}
\label{sigma:ac}
\sigma(\omega) = \frac{e^2n^*\tau_{\rm tr}^*}{2m_{c}^*[1+(\omega - \omega_{c}^*\eta)^2{\tau_{\rm tr}^*}^2\eta^2 ]},
\end{equation}
where parameters $m_{c}^*$, $\tau_{\rm tr}^*$ and $\omega_{c}^*$ denote corresponding quantities at 
a some (arbitrary) electron density $n = n^*$ and $\eta=\sqrt{n^*/n}$.
Equation~\eqref{sigma:ac} implies short-range scattering. The effect of electron density 
variation is demonstrated in Figs.~\ref{fig_9} and \ref{fig_9_15} where the photocurrent 
is plotted as a function of electron density for two values of magnetic field, 
$B=2$~T and $1.5$~T, respectively.  Note that all parameters used 
in the calculations are taken from experiments on magneto-transport and 
optical transmission performed for sample~\#4. 
One can see that the photocurrent as a function of electron density shows the pronounced resonance superimposed with quantum oscillations. Interestingly, for the linear electron dispersion $j_\mu$ contribution related with the difference of spin-up and spin-down subbands mobilities contains a smooth background caused by the dependence of the effective mass and scattering times on electron energy. 
Both figures show that the resonance position is close to the CR peak denoted as $n_{\rm CR}$ (see dotted lines) and shifts towards the smaller densities with a decrease of the $B$-field. This is in agreement with experimental data for sample~\#4 presented in Fig.~\ref{fig_6}
together with the absorption calculated after Eq.~(\ref{sigma:ac}) (see grey full lines). 
The dependence of $n_{\rm CR}$ on magnetic field calculated after Eq.~(\ref{sigma:ac})
agrees well with that obtained in the experiment, see Fig.~\ref{fig_6}(c).
Interestingly, Fig.~\ref{fig_6}(a,b) shows that while the SdH oscillations in the longitudinal resistance decrease and almost vanish with raising carrier density, the amplitude of the photocurrent oscillations gets strongly increased. This is just to the fact that the photocurrent is enhanced at the cyclotron resonance position, which in the sample~\#4 corresponds to rather high carrier density. 

To summarize this part, the photocurrent in unbiased HgTe-based QWs in the presence of magnetic field demonstrates $1/B$-periodic oscillations. The developed theory based on the spin-dependent asymmetric energy relaxation shows that the photocurrent oscillations stem from the crossing of Fermi level by Landau levels. The analysis shows that the photocurrent contains two contributions $j_n$ and $j_\mu$ caused by the electron spin polarization in the magnetic field and by the difference of spin-up and spin-down subbands mobilities. 

\begin{figure}[tb]
\includegraphics[width=\linewidth]{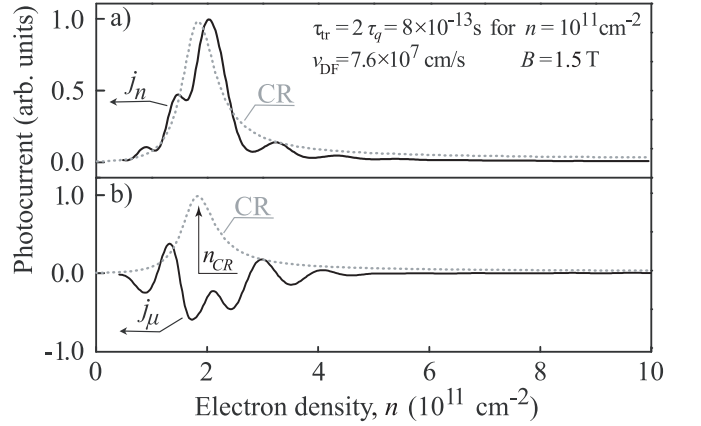}
\caption{Panels (a) and (b) show  photocurrent 
contributions $j_n$ and $j_\mu$ calculated
after Eqs.~\eqref{j1} and \eqref{j2}, respectively. 
Spin polarization and the high frequency mobilities 
in the opposite spin branches are calculated  for $B=1.5$~T
after~\eqref{Sz:D} and \eqref{mu:ac:D} applying
velocity of Dirac fermions $v_{\rm DF}=7.6\times 10^{7}$~cm/s close to that
experimentally determined  in Ref.~\onlinecite{Olbrich2013}.
In the calculations we assume the short-range scattering 
with $\tau_q= \tau_{\rm tr}/2 =4\times 10^{-13}$~sec for electron density $n=10^{11}$~cm$^{-2}$.
Dotted lines show CR  profile as 
a function of electron density. The radiation absorption 
is calculated after Eq.~\eqref{sigma:ac} for 
$B=1.5$~T. All parameters used in the calculations are taken from 
experiments on magneto-transport and  optical transmission performed for the sample~\#4.
Note that the photocurrent contributions and CR profile are normalized to 
their maximum values. 
}
\label{fig_9_15}
\end{figure}

\subsection{Photoresistance (photoconductivity) effect}\label{sec:photocond}

Finally, for completeness, we address  the oscillations of the photoresistance 
and compare these results with the experiment.
For the analysis of the photoresistance $\Delta R_{xx}$  we will use the data for 8~nm QW sample~\#5 shown in Fig.~\ref{fig_3} (c).
The microscopic origin of the photoresistance is related with the electron gas heating caused by the absorption of THz 
radiation and a corresponding reduction of the electron mobility, for review see e.g.~\cite{Ganichevbook}. 
In the model where the electron gas is described by an effective temperature $T_e$ the heating process is governed by the energy balance equation: 
\begin{equation}
\label{balance}
\frac{k_B(T_e - T_l)}{\tau_\epsilon} = {2}|e|[\mu_+(\omega) n_+ + \mu_-(\omega)n_-] |E_0|^2,
\end{equation}
where $T_l$ is the lattice temperature and $\tau_\epsilon$ is the energy relaxation time. The photoconductance tensor is, by definition, given by
\begin{equation}
\label{photocond}
\sigma^{(ph)}_{\alpha\beta} {= \Delta(\sigma_{\alpha\beta})} \equiv\sigma_{\alpha\beta}(T=T_e) - \sigma_{\alpha\beta}(T=T_l),
\end{equation}
where $\sigma_{\alpha\beta}$ is the static conductivity, and the photoresistance signal measured experimentally is given by
\begin{equation}
\label{DeltaRxx}
\Delta R_{xx} \propto \Delta\left(\frac{\sigma_{xx}}{\sigma_{xx}^2+\sigma_{xy}^2}\right),
\end{equation}
$\Delta(\ldots)$ denotes the variation of some quantity with and without illumination.
We recall that, similarly to Eq.~\eqref{mu:ac}, the components of static conductivity also contain oscillatory contributions. Particularly, for quantum wells with parabolic electron dispersion,
\begin{equation}
\label{Rxx}
R_{xx} \propto 1+ 2 \mathcal T_1 \exp{\left(- \frac{\pi}{\omega_c \tau_q}\right)}\cos{\left( 2\pi \frac{E_F}{\hbar\omega_c} \right)} \cos{\left(\pi \frac{\Delta_Z}{\hbar\omega_c}\right)}.
\end{equation}
Hence, an increase of an electron gas effective temperature $T_e$ mainly results in the suppression of the oscillatory factor because coefficient $\mathcal T_1$ given by Eq.~\eqref{Tk} exponentially decreases with an increase of temperature. As a result, the pronounced oscillatory photoresistance appears. The envelope function of the photoresistance oscillations at $B \lesssim B_{\rm CR}$
approximately follows the cyclotron resonance lineshape, because the heating is most efficient exactly in the resonance and becomes weaker the larger detuning, $|\omega_c - \omega|$. However, the oscillations of the photoresistance signal do not decay as fast as cyclotron resonance for $B>|B_{\rm CR}|$ because the oscillatory contributions increase with an increase of magnetic field. This is illustrated in Fig.~\ref{fig_PC_theor} where panel (a) shows the oscillatory part of resistivity calculated for the parameters close to that of $L_w=8$~nm sample~\#5, while panel (b) represents the photoresistance signal. By contrast to the calculations of photocurrents presented above, here we took into account both first and second harmonics of SdH oscillations, see below. The calculated photoresistance signal 
is presented in Fig.~\ref{fig_3}(c) (blue solid curve) together with the measured photoresistance (red solid curve) and calculated CR radiation absorption  (dashed curve). 
Apart a difference in the amplitudes the calculations are in a reasonable agreement with 
experiment.  


At last but not at least we address the influence of the ratio between the Landau level separation and Zeeman splitting. In linear transport for $\Delta_Z/(\hbar\omega_c) \approx 1/2$ the first harmonic in quantum oscillations 
of static resistivity vanishes, see Eq.~\eqref{Rxx} and Ref.~\onlinecite{tarasenkoZ} for details. This is caused by the mismatch of Landau levels and spin sublevels. Hence, second harmonic in the resistivity oscillations described by 
\begin{equation}
\label{Rxx:2ndH}
\mathcal T_2 \exp{\left(- \frac{2\pi}{\omega_c \tau_q}\right)}\cos{\left( 4\pi \frac{E_F}{\hbar\omega_c}\right)} \cos{\left(2\pi \frac{\Delta_Z}{\hbar\omega_c}\right)}
\end{equation}
with $\mathcal T_2 = {4\pi^2 k_B T_e}{[\hbar\omega_c \sinh{(4\pi^2 k_B T_e/\hbar\omega_c)}]^{-1}}$ becomes important. It results also in the frequency doubling of the photoresistance signal as clearly seen in Fig.~\ref{fig_PC_theor}. This effect is essential to describe the experimental data both on photoresistance and on SdH oscillations. 
By contrast, for the  previously discussed photocurrent and the spin polarization oscillations  both first and second harmonic vanish simultaneously if $\Delta_Z/(\hbar\omega_c) \approx 1$.
It is because photocurrent is caused by the imbalance of electron fluxes in two spin branches, therefore its extrema are realized when the mismatch of the Zeeman and Landau levels is maximal.


%

To conclude this section we note that at high lattice temperatures where oscillatory corrections to the conductivity governed by factors $\mathcal T_1$ and $\mathcal T_2$ vanish, the photoresistance can be still caused by the electron gas heating and corresponding change of the classical conductivity, i.e. via temperature dependence of the scattering times. In this case the photoresistance signal as a function of magnetic field follows the Lorentzian shape of cyclotron resonance  depicted by dotted curves in Fig.~\ref{fig_PC_theor} in agreement with experimental results, Fig.~\ref{fig_2}(c).



\begin{figure}[tb]
\includegraphics[width=\linewidth]{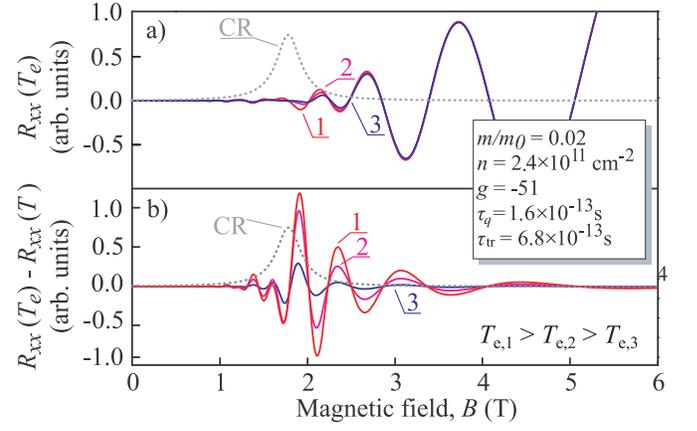}
\caption{
Oscillatory part of the resistivity (a) and 
the photoresistive signal (b) calculated after Eqs.~\eqref{photocond} -- \eqref{Rxx:2ndH}, respectively. 
Curves 1, 2, and 3 correspond to different electron 
temperatures, $T_e$, being $\propto \tau_\epsilon |E_0|^2$, see Eq.~\eqref{balance}.
The parameters are chosen close to those of 
8 nm  sample \# 5: $m=0.02m_0$, $n=2.4\times 10^{11}$~cm$^{-2}$, $g=-51$, 
$\tau_q=1.6\times 10^{-13}$~sec, momentum relaxation time 
$\tau_{\rm tr}=6.8\times 10^{-13}$~sec. Both first and second 
order oscillatory contributions are taken into account
and additional phase-shift $\varphi = -\pi/5$ was included to match the phase of experimental data in Fig.~\ref{fig_3}.
Dotted lines  show calculated cyclotron resonance absorption.
The data are given in arbitrary units.
}
\label{fig_PC_theor} 
\end{figure}

\section{Conclusions}

To conclude, the detailed experimental and theoretical studies of quantum magneto-oscillations of photocurrent in HgTe-based quantum wells have been performed. It has been demonstrated that $1/B$-periodic oscillations stem from the consecutive crossing of Fermi level by Landau levels and they become strongly enhanced at the cyclotron resonance conditions. The oscillations have been observed as a function of magnetic field for all three types of electron dispersion realized in HgTe quantum wells: normal parabolic, inverted parabolic, and linear. The developed theory explains photocurrent formation as a result of the spin-dependent asymmetric energy relaxation. It demonstrates that the photocurrent contains two contributions resulting from: (i) the magnetic field induced electron spin polarization and (ii) the difference of the electron mobilities in the spin-up and spin-down subbands in the presence of magnetic field, both of which contain oscillatory component.  The theory describes well all main experimental features.

\acknowledgements

Authors thank B. McCombe and S.A. Tarasenko for discussions.
The  support from the DFG (SPP 1666), 
Elite Network of Bavaria (K-NW-2013-247), RFBR and Russian President grants NSh-5062.2014.2, NSh-1085.2014.2
is gratefully acknowledged.


\begin{thebibliography}{99}
\bibitem{footnote013}
We note 
that studied (013)-grown QWs belong to the $C_1$ point group. The general form of the 
spin-dependent matrix element is 
$V_{\bm k'\bm k} = V_0 + \sum_{\alpha\beta} V_{\alpha\beta}\sigma_\alpha(k_\beta+k'_\beta)$ 
and the symmetry does not impose any restrictions on 
the components of second-rank pseudotensor $V_{\alpha\beta}$. Consequently, $\sigma_z k_y$ 
terms become possible 
and $y$ is some axis in the quantum well plane, whose direction may vary with the 
temperature, magnetic field, etc.
\bibitem{footnote_ee} At low temperatures the efficiency of electron-electron collisions scales as $(k_BT/E_F)^2$,~see Ref.~\onlinecite{chaplik71eng, hodges71, giuliani81}, where $E_F$ is the Fermi energy, while the temperature-induced suppression of quantum oscillations is governed by the factor $k_B T/\hbar\omega_c$.
\bibitem{footnote_xi} It is proportional to the product $\xi \sin{2\theta}$, where $\xi$ is the scattering asymmetry parameter introduced in Ref.~\onlinecite{Olbrich2013} and $\theta$ is the angle between [001] and [013] axes.
\bibitem{footnote_super}
Additional (radiative) broadening of the CR discussed in Refs.~\onlinecite{mikhailov,kono} is neglected hereafter.
\bibitem{misprint}Equations~\eqref{f1} and \eqref{mu:ac:D} correct the misprints which are present in the corresponding equations of Ref.~\onlinecite{briskot}.
\bibitem{footnote_Dyakonov}
We note, that the photovoltaic effect in field-effect transistors is also known to demonstrate quantum $1/B$-periodic magnetooscillations enhanced at the CR,\cite{Lifshits,Knap:exp1,Knap:exp2} however, it is described by the different microscopic mechanism related with the plasma waves excitation in 
gated
structures and, by contrast to $j_n$, is insensitive to the spin polarization.
\bibitem{footnote_smooth}
For the long-range scattering potential where $\tau_q \ne \tau_{\rm tr}$ additional contributions to $j_\mu$ may arise due to the energy dependence of the scattering times which is neglected in our consideration. These contributions may only provide a smooth background and are small by the parameter $\Delta_Z/E_F$.
\end{thebibliography}
\end{document}